\documentclass[prl,superscriptaddress,twocolumn,showpacs,preprintnumbers,amsmath,amssymb]{revtex4}

\usepackage{atlasphysics}
\usepackage{subfigure}
\usepackage{epsfig}
\usepackage{dcolumn}
\usepackage{bm}
\usepackage{graphicx}
\usepackage{xspace}
\usepackage{collref}
\usepackage[hyperindex,breaklinks]{hyperref} 

\def\simge{\mathrel{%
   \rlap{\raise 0.511ex \hbox{$>$}}{\lower 0.511ex \hbox{$\sim$}}}}
\def\simle{\mathrel{
   \rlap{\raise 0.511ex \hbox{$<$}}{\lower 0.511ex \hbox{$\sim$}}}}

\newcommand{\singletop}{\ensuremath{tW/tb/tqb}\xspace}
\newcommand{\otherdbn}{\ensuremath{WZ/ZZ/W\gamma}\xspace}
\newcommand{\ggF}{\ensuremath{gg\rightarrow H}\xspace}

\newcommand{\metrel}{\ensuremath{E_{\mathrm{T,rel}}^{\mathrm{\mathrm{miss}}}}\xspace}

\newcommand{\dphill}{\Delta\phi_{\ell\ell}}
\newcommand{\mT}{m_{\rm T}}

\newcommand{\mcatnlo}{{\sc MC@NLO}}
\newcommand{\ggtoww}{{\sc gg2ww}}
\newcommand{\alpgen}{{\sc Alpgen}}
\newcommand{\acermc}{{\sc AcerMC}}
\newcommand{\madgraph}{{\sc MadGraph}}
\newcommand{\powheg}{{\sc Powheg}}
\newcommand{\pythia}{{\sc Pythia}}
\newcommand{\herwig}{{\sc Herwig}}

\usepackage{mciteplus}

\begin{document}

\title{
{\vspace{-3.0 cm}\flushleft{\normalsize\normalfont{CERN-PH-EP-2011-190}}\flushright{\vspace{-0.86cm}\normalsize\normalfont{Submitted to Physical Review Letters}}\\[1.0cm]}
Search for the Higgs boson in the $H\rightarrow WW^{(*)}\rightarrow\ell^{+}\nu\ell^{-}\overline{\nu}$ 
  decay channel \\ in $pp$ collisions at $\boldmath{\sqrt{s}}$ = 7\,TeV with the ATLAS detector}
\author{The ATLAS Collaboration}

\begin{abstract} 

A search for the Higgs boson has been performed in the $H\rightarrow WW^{(*)}\rightarrow\ell^{+}\nu\ell^{-}\overline{\nu}$ 
channel ($\ell=e/\mu$) with an integrated luminosity of 2.05 ${\rm fb}^{-1}$ of $pp$ collisions 
at $\sqrt{s} = 7~{\rm TeV}$ collected with the ATLAS detector at the 
Large Hadron Collider. No significant excess of events over the
expected background is observed and limits on the Higgs boson production cross
section are derived for a Higgs boson mass in the range $110~{\rm GeV}<m_{\mathrm H}<300$~GeV.
The observations exclude the presence of a Standard Model Higgs boson with a mass 
$145<m_{\mathrm H}<206$~GeV at 95\% confidence level.
\end{abstract}


\maketitle

\normalsize

The Standard Model of particle physics postulates the existence of a complex scalar doublet with a vacuum expectation value, which
spontaneously breaks the electroweak symmetry, 
gives masses to all the massive elementary particles in the theory, and gives rise to a physical scalar known as the Higgs boson~\cite{prl_13_321,*prl_13_508,*prl_13_585}.
At the LHC, the Higgs boson is expected to be produced mainly 
through gluon fusion ($gg\rightarrow H$)~\cite{Harlander:2002wh,*Anastasiou:2002yz,*Ravindran:2003um,*Catani:2003zt,*Anastasiou:2008tj,*deFlorian:2009hc,*Aglietti:2004nj,*Actis:2008ug} due to the large gluon density,
although vector boson fusion ($qq\rightarrow qqH$)~\cite{Bolzoni:2010xr,*Ciccolini:2007jr,*Ciccolini:2007ec} is also important. Associated production of Higgs bosons
($WH$, $ZH$) also contributes more than 4\% to the total rate for $\mH \le 135$~GeV~\cite{LHCHiggsCrossSectionWorkingGroup:2011ti}.  
For $m_H \simge 135~\GeV$, $H\rightarrow WW^{(*)}$ is the dominant decay mode of the Higgs boson.  
Direct searches at LEP and the Tevatron exclude a Standard Model Higgs boson with a mass $m_{\mathrm H} < 114.4$~GeV 
or 156~GeV$<m_{\mathrm H} < 177$~GeV~\cite{Barate:2003sz,*tevhiggs} at 95\% confidence level (CL). 
The search for $H\rightarrow ZZ\rightarrow\ell\ell\nu\nu$ at ATLAS excludes a Standard Model Higgs boson with a mass $340<\mH<450$~\GeV, 
while the search for $H\rightarrow ZZ\rightarrow 4\ell$ excludes $191<\mH<197$~\GeV, $199<\mH<200$~\GeV, and $214<\mH<224$~\GeV~\cite{PhysRevLett.107.221802,*Aad2011435}.

\begin{table*}[t]
\begin{center}
\caption{
The expected numbers of signal ($m_{\mathrm H}=150$~\GeV) and background events after the requirements listed in the first column,
as well as the observed numbers of events in data.
All numbers are summed over lepton flavor.
}
\label{cutflow}
%\vspace*{0.2cm}
\resizebox{\textwidth}{!}{
\begin{tabular}{l|r@{$\,\pm \,$}l r@{$\,\pm \,$}l r@{$\,\pm \,$}l r@{$\,\pm \,$}l r@{$\,\pm \,$}l r@{$\,\pm \,$}l r@{$\,\pm \,$}l r@{$\,\pm \,$}l |c}
\hline
\hline
    \parbox{0pt}{~\\\vspace{0.45cm}}      $H+0$-jet Channel & \multicolumn{2}{c}{Signal} & \multicolumn{2}{c}{$WW$} & \multicolumn{2}{c}{$W+\mathrm{jets}$} & \multicolumn{2}{c}{$Z/\gamma^*+\mathrm{jets}$} & \multicolumn{2}{c}{$t\bar{t}$} & \multicolumn{2}{c}{$tW/tb/tqb$} & \multicolumn{2}{c}{$WZ/ZZ/W\gamma$} & \multicolumn{2}{c}{Total Bkg.} & Observed \\
\hline
Jet Veto                                 & 99 & 21     & 524 & 52   & 84 & 41   & 174 & 169   & 42  & 14   & 32  & 8   & 15  & 4   & 872 & 182   & 920\\
$p_{\rm T}^{\ell\ell}>30$ GeV            & 95 & 20     & 467 & 45   & 69 & 34   & 30  & 12    & 39  & 14   & 29  & 8   & 13  & 4   & 648 & 60    & 700\\
$m_{\ell\ell} < 50$ GeV                  & 68 & 15     & 118 & 15   & 21 & 8    & 13  & 8     & 7   &  4   & 5.8 & 1.8 & 1.9 & 0.6 & 166 & 19    & 199\\
$\Delta\phi_{\ell\ell} < 1.3$            & 58 & 13     & 91  & 12   & 12 & 5    & 9   & 6     & 6   &  3   & 5.8 & 1.8 & 1.7 & 0.6 & 125 & 15    & 149\\
$0.75\,m_{H}<m_{\rm T}<m_{H}$            & 40 & 9      & 52  & 7    & 5  & 2    & 2   & 4     & 2.4 &1.6   & 1.5 & 1.0 & 1.1 & 0.5 &  63 & 9     &  81\\
\hline
     \parbox{0pt}{~\\\vspace{0.45cm}}     $H+1$-jet Channel & \multicolumn{2}{c}{Signal} & \multicolumn{2}{c}{$WW$} & \multicolumn{2}{c}{$W+\mathrm{jets}$} & \multicolumn{2}{c}{$Z/\gamma^*+\mathrm{jets}$} & \multicolumn{2}{c}{$t\bar{t}$} & \multicolumn{2}{c}{$tW/tb/tqb$} & \multicolumn{2}{c}{$WZ/ZZ/W\gamma$} & \multicolumn{2}{c}{Total Bkg.} & Observed \\
\hline
1 jet                                    & 50 & 9      & 193 & 20   & 38 & 21   & 74  & 65    & 473 & 124  & 174 & 26  & 14  & 2    & 967 & 145   & 952\\
$b$-jet veto                             & 48 & 9      & 188 & 19   & 35 & 19   & 73  & 61    & 174 & 49   &  66 & 11  & 14  & 2    & 549 &  83   & 564\\
$|{\bf p}_{\rm T}^{\rm tot}| < 30$ GeV   & 39 & 7      & 154 & 16   & 18 &  9   & 38  & 32    & 106 & 30   &  50 &  9  & 9.7 & 1.5  & 376 &  48   & 405\\
$Z\rightarrow\tau\tau$ veto              & 39 & 7      & 150 & 17   & 18 &  8   & 34  & 23    & 102 & 23   &  48 &  8  & 9   & 2    & 361 &  38   & 388\\
$m_{\ell\ell} < 50$ GeV                  & 26 & 6      & 33  & 5    & 3.3&1.4   &  8  &  7    &  20 &  7   &  11 &  3  & 1.8 & 0.5  &  77 &  12   &  90\\
$\Delta\phi_{\ell\ell} < 1.3$            & 23 & 5      & 25  & 4    & 2.1&1.0   &  4  &  6    &  17 &  6   &   9 &  3  & 1.5 & 0.4  &  60 &  10   &  72\\
$0.75\,m_{H}<m_{\rm T}<m_{H}$            & 14 & 3      & 12  & 3    & 0.9&0.4   &1.3  & 1.9   &   8 &  2   & 4.0 & 1.6 & 0.7 & 0.3  &  28 &   4   &  29\\
\hline
\parbox{0pt}{~\\\vspace{0.45cm}}  Control Regions           & \multicolumn{2}{c}{Signal} & \multicolumn{2}{c}{$WW$} & \multicolumn{2}{c}{$W+\mathrm{jets}$} & \multicolumn{2}{c}{$Z/\gamma^*+\mathrm{jets}$} & \multicolumn{2}{c}{$t\bar{t}$} & \multicolumn{2}{c}{$tW/tb/tqb$} & \multicolumn{2}{c}{$WZ/ZZ/W\gamma$} & \multicolumn{2}{c}{Total Bkg.}           & Observed            \\
\hline
$WW$ 0-jet     ($m_H < 220$ GeV)         &1.7 & 0.4    & 223 & 30   & 20 & 15              &  6  &  8    &  25 & 10   &  15 & 4   &  8  & 3    & 296 &  36   & 296\\
$WW$ 0-jet     ($m_H \ge 220$ GeV)       & 10 & 2      & 173 & 23   & 24 & 12              & 13  & 19    &  15 &  6   &   8 & 3   & 3.3 & 0.6  & 236 &  33   & 258\\
$WW$ 1-jet     ($m_H < 220$ GeV)         &1.0 & 0.3    &  76 & 13   &  5 & 3               &  5  &  5    &  56 & 14   &  23 & 5   & 5.3 & 1.4  & 171 &  21   & 184\\
$WW$ 1-jet     ($m_H \ge 220$ GeV)       &5.8 & 1.5    &  51 &  9   &3.9 & 1.8             & 10  & 10    &  35 &  9   &  18 & 4   & 2.8 & 0.6  & 120 &  17   & 129\\
$\ttbar$ 1-jet                           &0.9 & 0.3    & 3.9 & 1.0  &\multicolumn{2}{c}{-} &  1  & 17    & 184 & 64   &  80 & 19  & 0.2 & 0.9  & 270 &  69   & 249\\
\hline
\hline
\end{tabular}
}
\end{center}
\vspace{-0.5cm}
\end{table*}

This Letter reports the results of a search for the Higgs boson in the channel 
$H\rightarrow WW^{(*)}\rightarrow\ell^{+}\nu\ell^{-}\overline{\nu}$~\cite{pr_55_167} 
($\ell=e/\mu$, but including contributions from $\tau\rightarrow e/\mu$ decays)
in 2.05 fb$^{-1}$ of $pp$ collisions at $\sqrt{s}=7$~TeV
recorded by the ATLAS detector during the LHC run of spring and summer 2011.
As described in detail below, the search examines events containing two leptons and up to one jet.  The main backgrounds are suppressed
by cuts on angular distributions, invariant masses, and $b$-jet tagging information.
The background normalization and composition is estimated in situ using several control samples defined by relaxing or reversing
selection cuts.
Similar searches were performed by CMS and ATLAS in 36~pb$^{-1}$~\cite{Chatrchyan201125} and  
35~pb$^{-1}$~\cite{Aad:2011qi}, respectively.
The ATLAS experiment~\cite{atlas} is a multipurpose particle
physics detector with forward-backward symmetric cylindrical geometry
allowing tracks within the pseudorapidity range $|\eta|<2.5$ and energy deposits in calorimeters covering $|\eta|<4.9$ to be reconstructed.
It is modeled using GEANT4~\cite{geant4,*ATLAS_simulation_paper} and simulated
events are reconstructed using the same software that is used to
perform the reconstruction on data.  The effects of multiple $pp$ interactions
(``in-time" pile-up) and residual energy deposits from neighboring bunch crossings (``out-of-time" pile-up) are modeled in the Monte Carlo (MC)
samples by superimposing a number of simulated minimum-bias events on the simulated signal and background
events.  MC samples with
different numbers of pile-up interactions are re-weighted to match the
conditions observed in the present data:  about 6 interactions per bunch crossing, with
a 50~ns bunch spacing.
The data used in this analysis were recorded during periods when
all ATLAS sub-detectors were operating under nominal conditions.
The events were triggered~\cite{Aad:2011xs} by requiring the
presence of a high-$p_{\rm T}$ electron or muon in the event.

Electron candidates are selected from clustered energy deposits in the
electromagnetic (EM) calorimeter with an associated track reconstructed in the inner detector and are required to satisfy a
stringent set of identification cuts~\cite{:2010yt} with
an efficiency of 71\% for electrons with transverse momentum
$E_{\rm T}>20$~GeV and $|\eta|<2.47$. 
Muons are reconstructed by combining tracks in the inner detector and
muon spectrometer. The efficiency of this reconstruction is $92$\%
for muons with $p_{\rm T}>20$~GeV and $|\eta|<2.4$.  
Events are required to have a primary vertex with $\ge 3$ tracks with $p_{\rm T}>0.4$~GeV.
For both electrons and muons, the track associated with the
lepton candidate is required to be consistent with having been produced
at the event's primary vertex.
Leptons are required to be isolated, satisfying stringent cuts on tracks and calorimeter depositions inside a 
cone $\Delta R = \sqrt{\Delta\phi^{2}+\Delta\eta^{2}} < 0.2$ around the lepton candidate, where $\Delta\phi$ and $\Delta\eta$ are the
transverse opening angle and pseudorapidity difference between the lepton and the track or energy deposit.
The lepton reconstruction 
efficiencies are evaluated with tag-and-probe methods using $Z\rightarrow\ell\ell$,
$J/\psi\rightarrow\ell\ell$, and $W\rightarrow\ell\nu$ events in data~\cite{Aad:2011mk}.

Jets are reconstructed from calibrated clusters using the
anti-$k_{t}$ algorithm~\cite{AntiKt} with radius parameter $R=0.4$.
Jet energies are calibrated using $E_{\rm T}$ and $\eta$ dependent 
correction factors based on MC simulation and validated by test 
beam and collision data studies \cite{jetEscaleNote}.
They are required to have
$E_{\rm T}>25$~GeV and $|\eta|<4.5$.  
Jets are identified as having been produced by $b$ quarks using an algorithm that combines
information about the impact parameter significance of tracks in the
jet and the topology of semi-leptonic $b$- and $c$-hadron decays~\cite{ATLAS-CONF-2011-102}.
The missing transverse momentum \MET~\cite{Aad:2011re} is reconstructed from calibrated
energy clusters in the calorimeters and the reconstructed momenta of the muons, 
which generally deposit only a small fraction of their energy in the calorimeters.
The $\MET$ distribution in the presence of pile-up has been studied, and both $\MET$ as a function of the number of reconstructed
primary vertices and $\MET$ as a function of the event's position in the bunch train are well-modeled by MC.

Exactly two opposite-sign lepton candidates ($e$ or $\mu$) with
$p_{\rm T}>15$~GeV for muons or $E_{\rm T}>20$~GeV for electrons are required.  The leading lepton must have transverse momentum $>25$~GeV so
the selected events have a high efficiency for the trigger selection.

After the selection of events with two leptons, the significant
backgrounds are the Drell-Yan process, $\ttbar$ and
single top (\singletop), $WW$, other diboson processes (\otherdbn),
and $W$+jets where a jet is misidentified as a lepton.
In addition to data-driven validations of the background estimates discussed later, 
MC simulations of the signal and backgrounds are studied in detail.
The \ggF and $qq\rightarrow qqH$ processes are modeled using
\powheg, with \pythia\ to handle the parton shower~\cite{Alioli:2008tz,*Nason:2009ai,*pythia}, 
and the \ggF Higgs boson \pt{} spectrum is reweighted to agree with the prediction of Ref.~\cite{Bozzi:2005wk}. 
\pythia\ is used to model $WH/ZH$ production.
Signal MC is generated in steps of 5~GeV for $m_{\mathrm H}$ below 200~GeV and in steps of 20~GeV for larger masses.
Signal expectations for intermediate mass values are obtained by linear interpolation of the signal efficiency.
The $\ttbar$, $s$-channel single top ($tb$), and  $qq/qg\rightarrow WW/WZ/ZZ$ processes are generated with
\mcatnlo, $t$-channel and $Wt$ single top with \acermc\ (interfaced to the parton shower algorithm in \pythia),
$gg\rightarrow WW$ with \ggtoww\ interfaced to the parton shower algorithm in \herwig~\cite{MCatNLO,*AcerMC,*gg2WW,*herwig},
$W\gamma$ with \madgraph\ interfaced to \pythia, and $W$+jets and $Z/\gamma^{*}$+jets
with \alpgen\ interfaced to \pythia~\cite{MadGraph,*alpgen}. 

If the two leptons have different flavors, their invariant mass ($m_{\ell\ell}$) is required to be above 10~GeV.
Otherwise, they must satisfy $m_{\ell\ell}>15$~GeV 
and they must lie outside the region with $|m_{\ell\ell}-m_{\mathrm Z}|<15$~\GeV\ to suppress backgrounds from $\Upsilon$ and $Z$ production, respectively.

The quantity $\metrel$ is defined as $\met$ if the angle $\Delta\phi$ between
the missing transverse momentum and the transverse momentum of the nearest lepton or jet
is greater than $\pi/2$, or $\met\sin(\Delta\phi)$ otherwise.
$\metrel$ is less sensitive to the mismeasurement of a single lepton or jet than $\met$.
To suppress backgrounds from multijet events and Drell-Yan production, 
it is required that $\metrel > 40$~\GeV ~if the two leptons have the same flavor, 
or $\metrel>25$~\GeV ~if they have different flavor. 

After these requirements, the data are separated into $H+0$-jet and $H+1$-jet~\cite{Mellado:2007fb} samples based on whether they have zero 
or exactly one jet.
In the $H+0$-jet channel, the dilepton system is required to have a large transverse boost, $p_{\rm T}^{\ell\ell}>30$~\GeV, to suppress backgrounds from
$Z$+jets and continuum $WW$ production.

\begin{figure*}[htb]
  \begin{center}
  \begin{tabular}{lccr}
  \includegraphics[width=0.55\columnwidth]{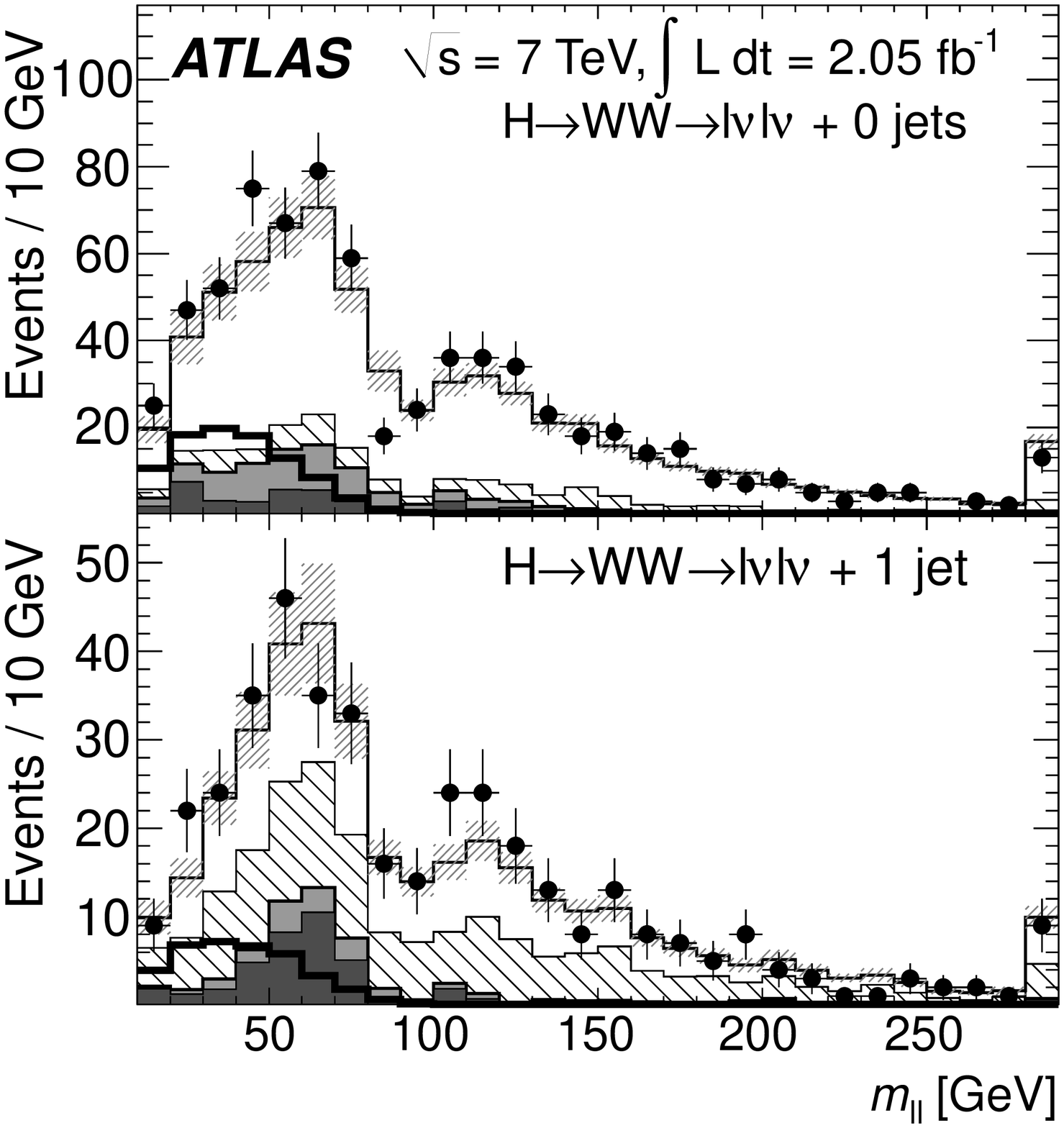} &
  \includegraphics[width=0.55\columnwidth]{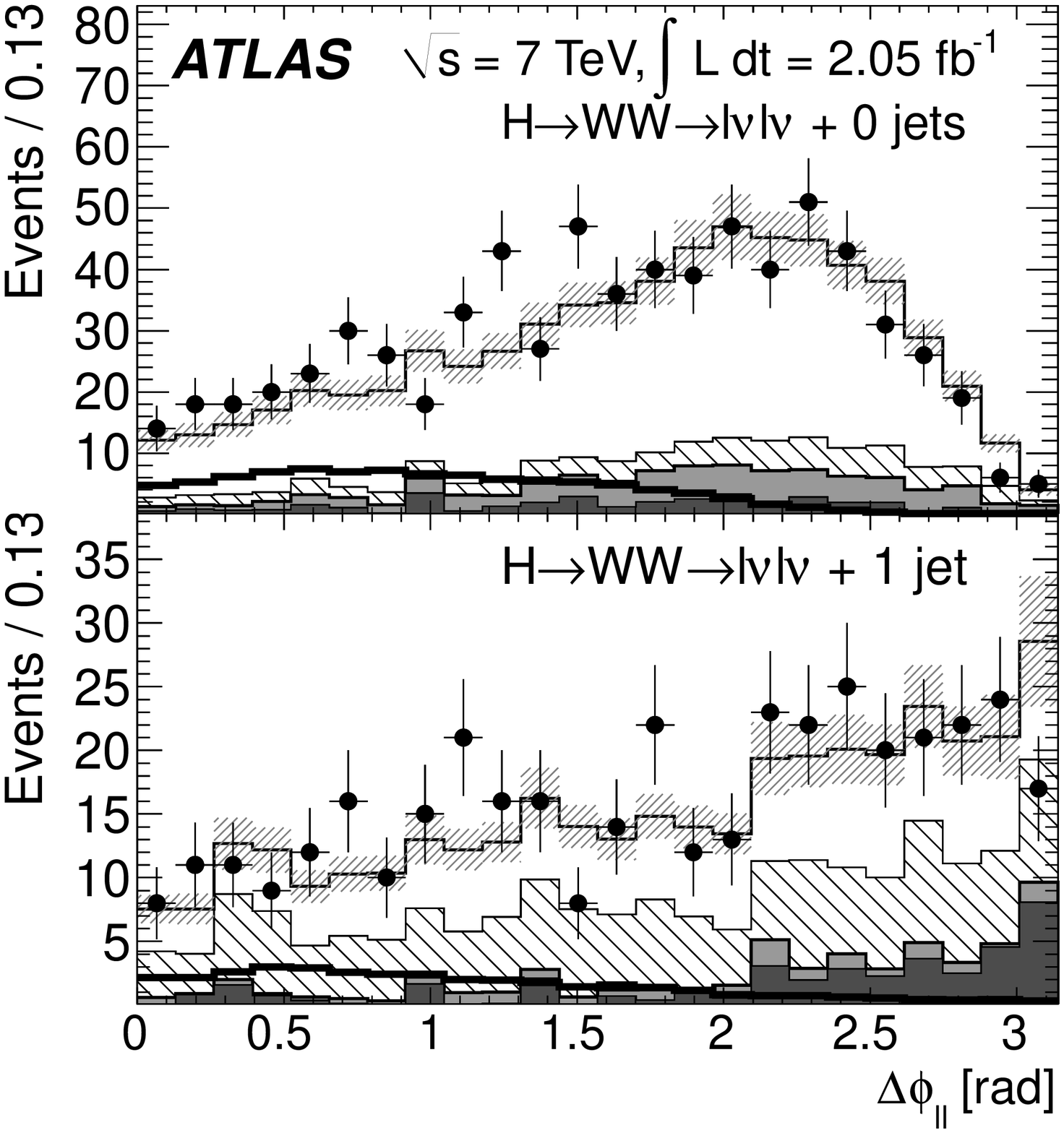} &
  \includegraphics[width=0.55\columnwidth]{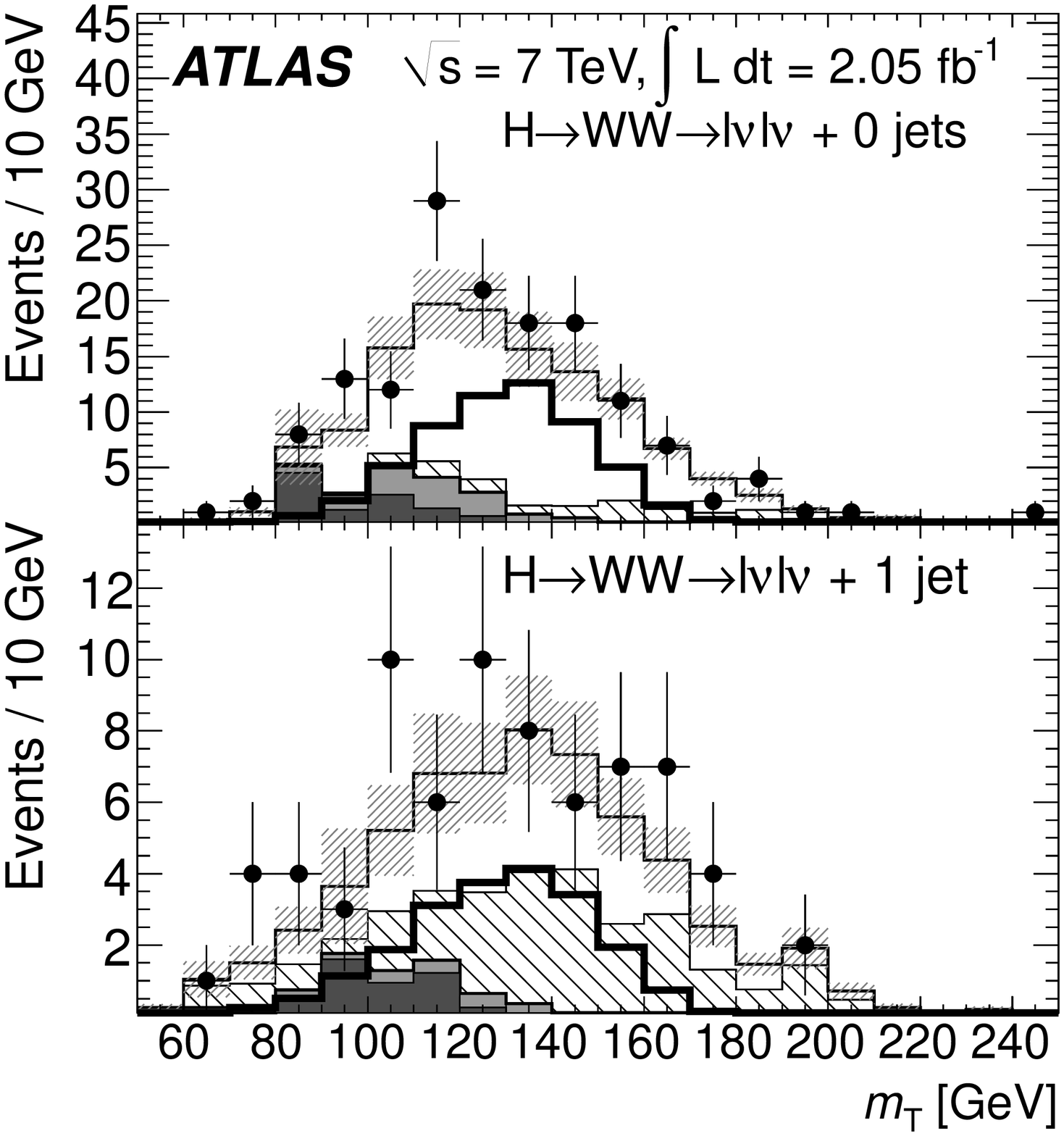} &
  \includegraphics[width=0.30\columnwidth]{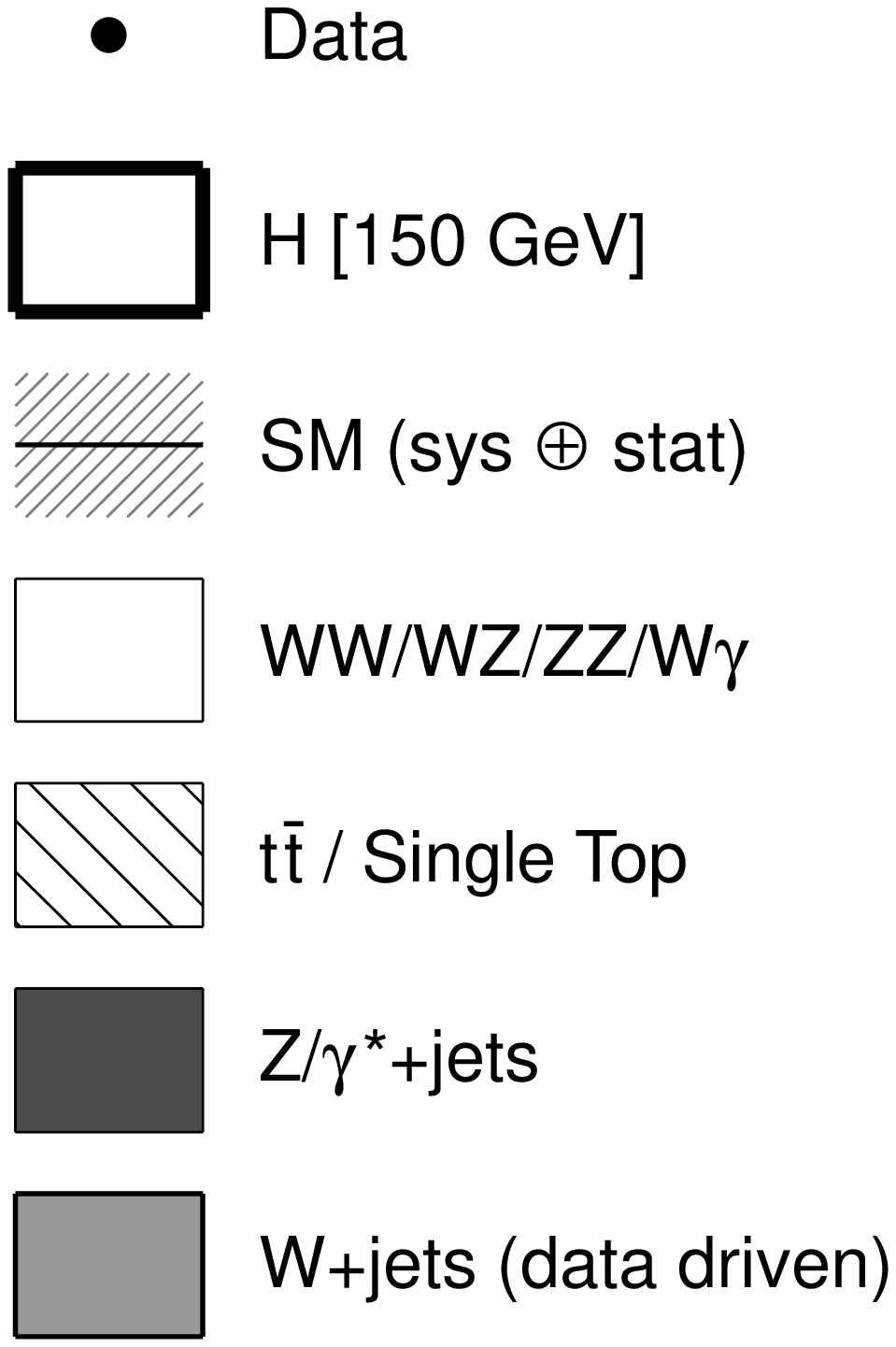} \\
  \end{tabular}
  \end{center}
  \vspace{-0.5cm}
  \caption{Distributions of $m_{\ell\ell}$ (left), $\Delta\phi_{\ell\ell}$ (center), and $\mT$ (right).
    The top row shows the selection for the $H+0$-jet channel and the bottom row for the $H+1$-jet channel.
    The left and central plots are shown after the $p_{\rm T}^{\ell\ell}$ selection for the $H+0$-jet channel and
    after the $|{\bf p}_{\rm T}^{\rm tot}|$ cut for the $H+1$-jet channel.
    For the rightmost plots, the distributions are shown after all the cuts for $m_{H}=150$~\GeV\ except the cut on $\mT$ itself.
    The background distributions are stacked, so that the top of the
    diboson background coincides with the Standard Model (SM) line which includes the
    statistical and systematic uncertainties on the expectation in the
    absence of a signal.
    The expected signal for $m_H$=150~GeV is shown as a separate thicker line, and the final bin includes the overflow.
  \label{fig:datamc}}
  \vspace{-0.2cm}
\end{figure*}

To suppress background from top-quark production,
events in the $H+1$-jet channel are rejected if the jet is identified as the decay of a $b$-quark.
These candidates are further required to have $|{\bf p}_{\rm T}^{\rm tot}|<30$~\GeV, where ${\bf p}_{\rm T}^{\rm tot}$ is the vector sum of the transverse momenta of the jet, the two leptons, and the $\met$ vector.
This latter selection suppresses events with significant hadronic activity that recoils against the ${\bf p}_{\rm T}^{\rm tot}$ system but 
does not leave high \pT\ jets in the detector.
In the $H+1$-jet channel, the event is required to 
pass the $Z\rightarrow\tau\tau$ rejection cut used in the $H\rightarrow WW$ analysis of Ref.~\cite{Asai:2004ws}.

Top and $WW$ backgrounds are suppressed by an upper bound on $m_{\ell\ell}$.  Because the $m_{\ell\ell}$ distribution for the signal depends strongly on
$m_{\mathrm H}$, the chosen upper bound depends on the Higgs boson mass hypothesis.  For $m_{\mathrm H}<170$~\GeV, $m_{\ell\ell}<50$~\GeV ~is
required, while for $170\leq m_{\mathrm H}<220$~\GeV, the cut is $m_{\ell\ell}<65$~\GeV.  For $m_{\mathrm H}\geq 220$~GeV, the requirement is $50<m_{\ell\ell}<180$~\GeV.

For $m_{\mathrm H}<220$~GeV, an upper bound is imposed on the azimuthal angle between the two
leptons to exploit differences in spin correlations between signal and background: $\dphill< 1.3$ for $m_{\mathrm H}<170$~\GeV, or $\dphill<1.8$ for $m_{\mathrm H}<220$~GeV.
The final requirement uses the transverse mass $\mT$~\cite{Barr:2009mx} which is defined 
as $(\mT)^{2}=m_{v}^{2} + 2(e_{v}|{\bf p}_{{\rm T},i}|-{\bf p}_{{\rm T},v}\cdot {\bf p}_{{\rm T},i})$, where the subscripts $v$ and $i$ denote the visible and invisible decay products
and $e_{v}=\sqrt{{\bf p}_{{\rm T},v}\cdot {\bf p}_{{\rm T},v} + m_{v}^{2}}$ denotes the transverse energy.
The transverse mass $\mT$ is required to lie within $0.75 m_{\mathrm H} < \mT < m_{\mathrm H}$ if $m_{\mathrm H}<220$~\GeV\ or
$0.6 m_{\mathrm H} < \mT < m_{\mathrm H}$\ otherwise.  The upper bound on this window reduces the $WW$ and top backgrounds and excludes regions of phase space where interference effects between 
the signal and the $gg\rightarrow WW$ background are large~\cite{Campbell:2011cu}. 

Table~\ref{cutflow} shows the expected and observed event yields after these cuts.
As described below, the $W$+jets background is entirely determined from data, whereas for the other processes 
the expectations are based on simulation, with $Z/\gamma^*+\mathrm{jets}$, $t\bar{t}$, and $tW/tb/tqb$ corrected 
by scale factors derived from control samples.  The uncertainties shown are the sum in quadrature of 
systematic uncertainties and statistical errors due to the finite number of MC events.
Figure~\ref{fig:datamc} shows the
distributions of $m_{\ell\ell}$ and $\Delta\phi_{\ell\ell}$ before the final cut on $m_{\ell\ell}$, and the distribution of $\mT$ after the cut on $\Delta\phi_{\ell\ell}$.

The background from $W$+jets events where one jet is misidentified as a lepton is estimated from data using a control sample where
one of the two leptons satisfies a loosened set of identification and isolation criteria but not the full set of criteria normally used.
The extrapolation from this control sample to the signal region is extracted from dijet events~\cite{Aad:2011kk}. 

The Drell-Yan background is corrected for
mis-modeling of the distribution of $\met$ at high values
based on the observed difference between the 
fraction of events passing the $\metrel > 40$~GeV selection
in data and MC simulation for events with $m_{\ell\ell}$ within 10~\GeV\ of the $Z$ boson mass. 
The correction factors are all found to be between 0.8 and 0.9, which indicates that the background in the signal
region is about 15\% less than the MC estimates.

The $WW$ and top backgrounds are normalized by a simultaneous fit to the numbers of observed events in the signal region and several control samples.
A sample enriched in $WW$ background is defined by removing the selections on $\mT$ and $\dphill$ and changing the selection
on $m_{\ell\ell}$.  For $m_{\mathrm H}<220$~\GeV, the cut is changed to $m_{\ell\ell}>80$~\GeV, while for $m_{\mathrm H}>220$~\GeV, the control region is the
union of the regions with $15<m_{\ell\ell}<50$~\GeV\ and $m_{\ell\ell}>180$~\GeV.  This control sample is studied separately for the $H+0$-jet channel
and the $H+1$-jet channel, and the observed yields are consistent with expectations in both cases.  The yields in these control regions,
shown in Table~\ref{cutflow}, are 
propagated to the signal region using scale factors computed with MC.

In the $H+0$-jet channel, the top-enriched control sample consists of the same preselected sample used in the rest of this
analysis:  events with two leptons and $\metrel$.  The scale factor used to propagate the $\ttbar$ yield from this 
sample to the signal region is estimated as the square of the efficiency for one top decay to survive the jet veto
(estimated using another control sample, defined by the presence of an additional $b$-jet), with a correction computed using MC to account for the presence of single top~\cite{PhysRevD.84.096005}.
A sample enriched in top background is defined for the $H+1$-jet channel by reversing the $b$-jet veto and removing the cuts on
$\dphill$, $m_{\ell\ell}$, and $\mT$.  The extrapolation to the signal region is done using a scale factor computed using MC.  
The control samples for top in the $H+0$-jet and $H+1$-jet channels also normalize the top contamination in the corresponding $WW$ control regions.
In both cases, the estimated top backgrounds are consistent with the expected yields in Table~\ref{cutflow}.

The signal significance and limits on Higgs boson production are derived from a 
likelihood function that is the product of the Poisson
probabilities of each of the lepton flavor and jet multiplicity
yields for the signal selections,  the $WW$+0-jet and $WW$+1-jet control regions, and
top control region for the $H+1$-jet channel. 
The normalization of the
signal, the $WW$ cross sections for the $H+0$-jet and $H+1$-jet channels, and the top cross section for the $H+1$-jet channel
are allowed to vary independently; the control regions included in the fit constrain all of these except the signal yield.  All other components are 
normalized to their expectations scaled by nuisance parameters constrained by Gaussian terms that include the systematic uncertainties described below.
The results from the control sample measurements for the top background in the $H+0$-jet channel and for the $W$+jets and Drell-Yan backgrounds everywhere are used as the expected values
for the corresponding backgrounds in the fit.  Since these contributions are small, the control samples themselves are not explicitly modeled in the fit
as they are for top in the $H+1$-jet channel and for $WW$ everywhere.

The systematic uncertainties include contributions from the 3.7\% uncertainty in the luminosity~\cite{Aad:2011dr,*ATLAS-CONF-2011-116}, 
and from theoretical uncertainties, which are  -8/+12\% and $\pm$8\% from 
the QCD scale and 1\% and 4\% from the parton density functions, for $gg\rightarrow H$ and $qq\rightarrow qqH$ respectively.
Additional theoretical uncertainties on the acceptance are assessed as described in Ref.~\cite{ATL-PHYS-PUB-2011-011}.
In particular, the uncertainty in the assignment of events to jet multiplicity bins is included separately as an uncertainty on the cross section of each bin, 
calculated from the approximate 10\% and 20\% uncertainties of the inclusive 0-jet and 1-jet cross sections,
respectively.

Several sources of measurement uncertainty are taken into account.
The uncertainty on the jet energy scale is less than 10\% on the global scale including flavor composition effects,
with an additional uncertainty of up to 7\% due to pile-up~\cite{jetEscaleNote}.
The electron and muon efficiencies are determined from samples of $W$ and $Z$ boson data 
with uncertainties of 2-5\% and 0.3-1\%, respectively, depending on $|\eta|$ and $p_{\rm T}$.
Uncertainties are $<1$\% and $<0.1$\%, respectively, on the lepton energy scale and $<0.6$\% and $<5$\% on the resolution~\cite{Aad:2011mk}.
The uncertainties on the $b$-tagging efficiency and mistag rate are 6-15\% and up to 21\%, respectively~\cite{ATLAS-CONF-2011-102}.
A 13\% uncertainty is applied to the energy scale for low-$p_{\rm T}$ depositions 
in the $\met$ measurement.
All these sources of detector uncertainty 
are propagated to the result by varying
reconstructed quantities and observing the effect on the expected yields.
For the $WW$ background, the total (theoretical and experimental) uncertainty on the ratio of cross sections in the signal and control regions
is 7.6\% in the $H+0$-jet channel and 21\% in the $H+1$-jet channel; 
for the top background in $H+1$-jet the total for the extrapolation to the signal region is 38\%, and 29\% to the $WW$ control region.

\begin{figure}
\begin{center}
\includegraphics[width=0.48\textwidth]{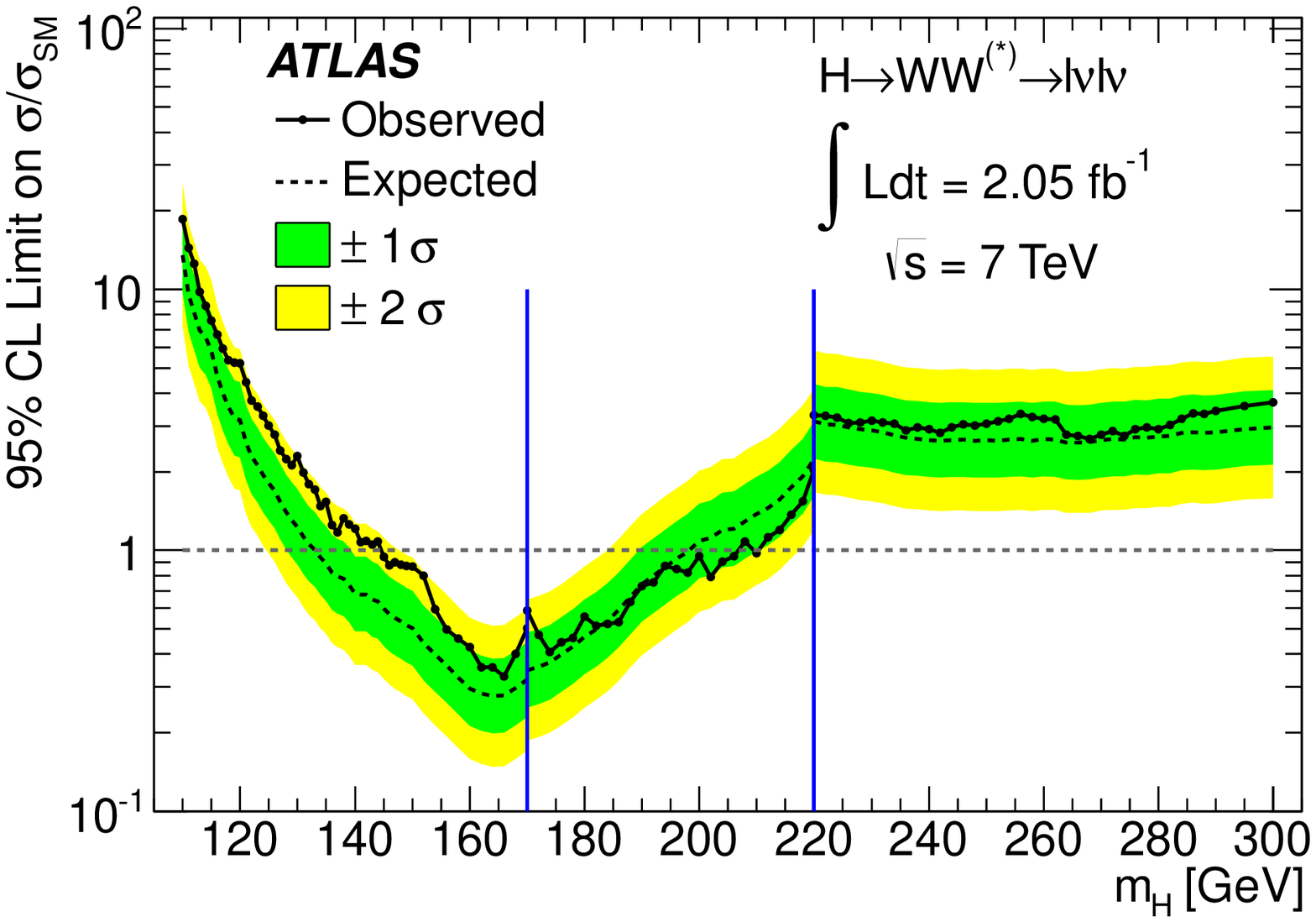}
\vspace*{-0.5cm}
\caption{The expected (dashed) and observed (solid) 95\% CL upper limits on the cross section,
normalized to the Standard Model cross section, as a function of the Higgs boson mass.  Expected limits are
given for the scenario where there is no signal.  The vertical lines in the
curves indicate the points where the selection cuts change, and the bands around the dashed line indicate the expected statistical fluctuations
of the limit.
}\label{limits}
\vspace*{-0.7cm}
\end{center}
\end{figure}

No significant excess of events is observed.  The largest observed deviation from the expected background is 1.9$\sigma$.
A 95\% CL upper bound is set on the Higgs boson cross section 
as a function of $m_{\mathrm H}$ using the $CL_{s}$ formalism~\cite{0954-3899-28-10-313,*2011EPJC...71.1554C}.  
Figure~\ref{limits} shows the expected and observed limits.
Discontinuities occur where the selection changes, since the signal regions there are less statistically correlated between adjacent masses.  
In the absence of a signal, one would expect to exclude a Standard Model Higgs boson in the range $134<m_{\mathrm H}<200$~GeV at the 95\% CL.
The Higgs boson mass interval excluded by the measurements presented in this Letter, $145<m_{\mathrm H}<206$~GeV, 
is consistent with that expectation. 
This measurement excludes, at 95\% CL, 
a larger part of the mass range favored by the electroweak fits than previous limits~\cite{:2010vi}.

\section{Acknowledgments}

We thank CERN for the very successful operation of the LHC, as well as the
support staff from our institutions without whom ATLAS could not be
operated efficiently.

We acknowledge the support of ANPCyT, Argentina; YerPhI, Armenia; ARC,
Australia; BMWF, Austria; AHAS, Azerbaijan; SSTC, Belarus; CNPq and FAPESP,
Brazil; NSERC, NRC and CFI, Canada; CERN; CONICYT, Chile; CAS, MOST and
NSFC, China; COLCIENCIAS, Colombia; MSMT CR, MPO CR and VSC CR, Czech
Republic; DNRF, DNSRC and Lundbeck Foundation, Denmark; ARTEMIS, European
Union; IN2P3-CNRS, CEA-DSM/IRFU, France; GNAS, Georgia; BMBF, DFG, HGF, MPG
and AvH Foundation, Germany; GSRT, Greece; ISF, MINERVA, GIF, DIP and
Benoziyo Center, Israel; INFN, Italy; MEXT and JSPS, Japan; CNRST, Morocco;
FOM and NWO, Netherlands; RCN, Norway; MNiSW, Poland; GRICES and FCT,
Portugal; MERYS (MECTS), Romania; MES of Russia and ROSATOM, Russian
Federation; JINR; MSTD, Serbia; MSSR, Slovakia; ARRS and MVZT, Slovenia;
DST/NRF, South Africa; MICINN, Spain; SRC and Wallenberg Foundation,
Sweden; SER, SNSF and Cantons of Bern and Geneva, Switzerland; NSC, Taiwan;
TAEK, Turkey; STFC, the Royal Society and Leverhulme Trust, United Kingdom;
DOE and NSF, United States of America.

The crucial computing support from all WLCG partners is acknowledged
gratefully, in particular from CERN and the ATLAS Tier-1 facilities at
TRIUMF (Canada), NDGF (Denmark, Norway, Sweden), CC-IN2P3 (France),
KIT/GridKA (Germany), INFN-CNAF (Italy), NL-T1 (Netherlands), PIC (Spain),
ASGC (Taiwan), RAL (UK) and BNL (USA) and in the Tier-2 facilities
worldwide.

\bibliographystyle{apsrevM}
\bibliography{paper,Higgs_paper}
\onecolumngrid
\clearpage
% ATLAS Collaboration author list for 11-OCT-2011
% Data extracted on 16-Mar-2012 for paperid 146
%\documentclass[11pt]{article}
%\usepackage{a4wide}\begin{document}
\begin{flushleft}
{\Large The ATLAS Collaboration}

\bigskip

G.~Aad$^{\rm 48}$,
B.~Abbott$^{\rm 110}$,
J.~Abdallah$^{\rm 11}$,
A.A.~Abdelalim$^{\rm 49}$,
A.~Abdesselam$^{\rm 117}$,
O.~Abdinov$^{\rm 10}$,
B.~Abi$^{\rm 111}$,
M.~Abolins$^{\rm 87}$,
H.~Abramowicz$^{\rm 152}$,
H.~Abreu$^{\rm 114}$
% ...
\bigskip

% ...
$^{10}$ Institute of Physics, Azerbaijan Academy of Sciences, Baku, Azerbaijan\\
$^{11}$ Institut de F\'isica d'Altes Energies and Departament de F\'isica de la Universitat Aut\`onoma  de Barcelona and ICREA, Barcelona, Spain\\
% ...
$^{48}$ Fakult\"{a}t f\"{u}r Mathematik und Physik, Albert-Ludwigs-Universit\"{a}t, Freiburg i.Br., Germany\\
$^{49}$ Section de Physique, Universit\'e de Gen\`eve, Geneva, Switzerland\\
% ...
$^{87}$ Department of Physics and Astronomy, Michigan State University, East Lansing MI, United States of America\\
% ...
$^{110}$ Homer L. Dodge Department of Physics and Astronomy, University of Oklahoma, Norman OK, United States of America\\
$^{111}$ Department of Physics, Oklahoma State University, Stillwater OK, United States of America\\
% ...
$^{114}$ LAL, Univ. Paris-Sud and CNRS/IN2P3, Orsay, France\\
% ...
$^{117}$ Department of Physics, Oxford University, Oxford, United Kingdom\\
% ...
$^{152}$ Raymond and Beverly Sackler School of Physics and Astronomy, Tel Aviv University, Tel Aviv, Israel\\
% ...
\end{flushleft}

%\end{document}

\end{document}